\documentclass{emulateapj}
\usepackage{apjfonts}
\journalinfo{The Astrophysical Journal}
\shorttitle{The H\lowercase{e-sd}B CPD$-$20~1123}
\shortauthors{Vennes et al.}

\begin{document}

\title{CPD$-$20~1123 (Albus~1) is a bright H\lowercase{e}-B Subdwarf}

\author{St\'ephane Vennes\altaffilmark{1,3}, Ad\'ela Kawka\altaffilmark{2,3}, and J. Allyn Smith\altaffilmark{4}}

\email{svennes@fit.edu, kawka@sunstel.asu.cas.cz, smithj@apsu.edu.}

\altaffiltext{1}{Department of Physics and Space Sciences,
Florida Institute of Technology, Melbourne, FL 32901, USA.}
\altaffiltext{2}{Astronomick\'y \'ustav Akademie v\v{e}d {\v C}esk\'e republiky,
Fri{\v c}ova 298, CZ-251 65 Ond{\v r}ejov, Czech Republic.}
\altaffiltext{3}{Visiting astronomer, Cerro Tololo Inter-American Observatory, National Optical Astronomy Observatories, which are operated by theAssociation of Universities for Research in Astronomy, under contract with the National Science Foundation.}
\altaffiltext{4}{Department of Physics and Astronomy, Austin Peay State University, Clarksville, TN 37044  USA.}

\begin{abstract}

Based on photometric and astrometric data it has been  proposed that Albus~1 (also known as CPD$-$20~1123) might be a hot white dwarf similar to G191-B2B or, alternatively, a hot  subdwarf. We obtained a series of optical spectra showing that  CPD$-$20~1123 is a bright He-B subdwarf. We analyzed the \ion{H}{1} Balmer and \ion{He}{1} line spectra and measured $T_{\rm eff}=19800\pm400$ K, $\log{g}=4.55\pm0.10$, and $\log{N({\rm He})/N({\rm H})}=0.15\pm0.15$. This peculiar object belongs to a family of evolved helium-rich stars that may be the products of double-degenerate mergers, or, alternatively,  the products of post horizontal- or giant-branch evolution.

\end{abstract} 


\section{Introduction}

Albus 1 is a blue star ($B_T=11.75\pm0.07$ mag)  but its true nature remains to be established. It was first recorded as CPD$-$20~1123 \citep{gil1896} with a photographic magnitude $m_p=10.6$\,mag but it did not attract much attention until \cite{cab2007} proposed that the star might be a nearby ($d\sim40$ pc) hot white dwarf, or, alternatively, a hot subdwarf. They noted that the star stands out as the bluest of over a thousand objects investigated in an area of $\approx 18$ deg$^2$ and covered by the Tycho-2 and 2MASS catalogues. The object is also known as TYC~5940~962~1 \citep{hog2000} and 2MASS~J06061339$-$2021072 \citep{cut2003}. \citet{cab2007} located the star at 
${\rm R.A.} =06^{\rm h} 06^{\rm m} 13^{\rm s}.39,\ {\rm  decl.}=-20\arcdeg21\arcmin 07\arcsec.3$ (J2000).

We present in \S 2 a series of optical spectra of CPD$-$20~1123 obtained at the Cerro Tololo Inter-American Observatory (CTIO) which show the star to be a helium-rich B subdwarf. The star appears  very similar to another well studied peculiar subdwarf, PG~0229$+$064 \citep{gre1986, moe1990, ahm2003}, but with markedly stronger optical \ion{He}{1} lines.  Hot subdwarf B (sdB) stars are core helium burning stars that lie at the hot  end of the horizontal branch, i.e., the extreme horizontal branch stars. These stars cover a narrow mass range around $0.5$  M$_\odot$ and have very thin hydrogen ($< 0.02$ M$_\odot$) envelopes  \citep{heb1986}. They possibly evolved via two main branches.  The first is through single star evolution where the star fails to ascend  the asymptotic giant branch and loses most of its mass via extensive mass loss \citep{dcr1996}. The second is through binary star  evolution, with one proposed scenario involving the merger of two helium white  dwarfs \citep{ibe1990, sai2000}. \citet{han2002,han2003} conducted population syntheses of  sdB stars and showed that most should have evolved in binary systems.  Helium-rich subdwarfs  (He-sdB) have a higher abundance of helium
and a systematically lower surface gravity compared to most sdB stars \citep{ahm2003}. The origin of He-sdB stars is still uncertain, but \citet{ahm2004} consider that a white dwarf merger is the most likely case.  We complete our investigation of  CPD$-$20~1123 with a model atmosphere analysis in \S 3, and we conclude in \S 4.
 
\section{Observations}

We have obtained three spectra of CPD$-$20~1123 (Table~\ref{tbl1}) using the R-C Spectrograph attached to the CTIO 4.0 m Blanco Telescope. We employed the KPGL2 grating in the first order and centered at 5109 \AA\ resulting in a dispersion of  $1.992$ \AA\ per pixel. We also used the order-sorting filter WG360 which provided us with an effective coverage from $\lambda=3700$ to 7200 \AA. We observed at the parallactic angle and the slit width was set at 1.5$\arcsec$ which resulted in a resolution element 4 pixels wide and a FWHM$\approx8$ \AA. Finally we obtained a HeNeAr comparison arc after the series of exposures. We calibrated the spectrometer response with the flux standards EG 131 and Feige 110. However, the spectra of CPD$-$20~1123 were obtained at a high airmass under poor seeing conditions, and  with possible obstruction by clouds near the horizon which resulted in considerable light losses. Nonetheless, the relative flux spectrum  is  of good quality with a signal-to-noise ratio of $\approx150$, although we could not secure the absolute flux scale. All data were reduced using standard procedures within IRAF.\footnote{IRAF is distributed by the National Optical Astronomy Observatories, which are operated by the Association of Universities for Research in Astronomy, Inc., under cooperative agreement with the National Science Foundation.} 

\begin{deluxetable}{ccc}
\tablewidth{0pt}
\tablecaption{CTIO Observation Log \label{tbl1}}
\tablehead{
 \colhead{Date and time} & \colhead{Exposure time} & Airmass}
\startdata
 UT 2007 July 12 10:19 &  120 s  & 2.83  \\
 UT 2007 July 12 10:25 &  120 s   & 2.70 \\
 UT 2007 July 12 10:29 &  120 s  & 2.60   
\enddata
\end{deluxetable}

\begin{figure}
\plotone{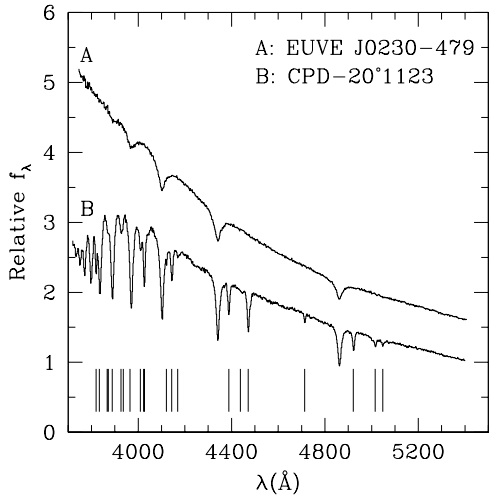}
\caption{Summed CTIO spectrum of CPD$-$20~1123 normalized at $\lambda=5500$\AA\ and a spectrum of the DA white dwarf  EUVE~J0230$-$479 also normazlied at 5500\AA\ but shifted up by 0.5 unit.  \ion{He}{1} lines are marked with vertical lines.\label{fig1}}
\end{figure}

Figure~\ref{fig1} shows the summed spectrum. The hydrogen Balmer line series and numerous \ion{He}{1} lines are evident. The \ion{He}{1} lines appear stronger than in ordinary B subdwarf or main-sequence stars. The object bears similarities with a sample of He-rich subdwarf B stars studied by \cite{ahm2003}.  The spectrum of CPD$-$20~1123 is also compared to a spectrum of the hot white dwarf EUVE~J0230$-$479 \citep[$T_{\rm eff}=64800$ K, $\log{g}=7.72$;][]{ven1996} obtained with the same set-up on 2007 July 14. The white dwarf EUVE~J0230$-$479 and the proto-typical DA G191-B2B share similar properties but Figure~\ref{fig1} shows that CPD$-$20~1123 is not a hot white dwarf.

\section{Analysis and Discussion}
 
\begin{figure}
\plotone{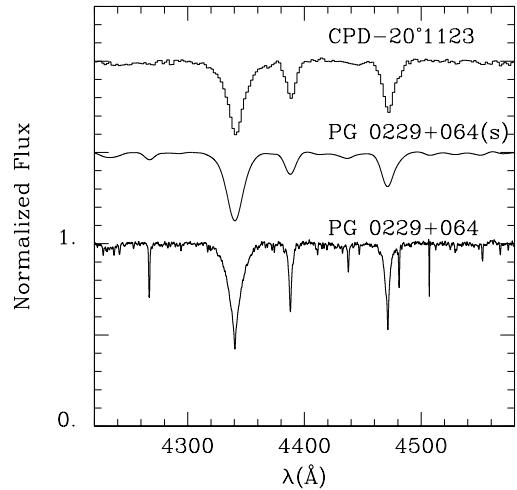}
\caption{Comparing  CPD$-$20~1123 (shifted up by 1.0 unit) and PG~0229$+$064. The spectrum of PG~0229$+$064 is presented at the nominal resolution of 0.8 \AA\  as well as degraded to a resolution of  8 \AA\ (labeled ``s" and shifted up by 0.5 unit).\label{fig2}}
\end{figure}

The peculiar helium line spectrum clearly indicates  an evolved star. Fig.~\ref{fig2} compares
CPD$-$20~1123 and PG~0229$+$064 \citep{gre1986, moe1990, ahm2003}. We obtained the spectrum of PG~0229$+$064 from the Isaac Newton Group Archive. The spectrum was originally obtained by \cite{ahm2003} using the ISIS double-beam spectrograph attached to the 4.2m William Herschel Telescope. The helium lines appear somewhat weaker in PG~0229$+$064 than in CPD$-$20~1123 which may indicate that the helium abundance is somewhat higher in the latter. Following a strict application of the subdwarf classification scheme proposed by \citet{dri2003} CPD$-$20~1123, much like PG~0229$+$064, would be listed as a sdB3V:He15 where a  classification index of He15 implies a deeper H$\gamma$ line over the \ion{He}{1}$\lambda4471$ line. The absence of \ion{He}{2} is consistent with a relatively low temperature compared to other helium-rich subdwarf  stars. In order to find out whether or not the atmosphere is helium-rich it is necessary to perform a model atmosphere analysis.

We computed a series of non-LTE model atmospheres using  TLUSTY version 200 and 
SYNSPEC version 48 \citep{hub1995}. The three-dimensional grid points are located at  $T_{\rm eff}=18000, 20000$, and 22000 K, at $\log{g}=4.5, 5.0$ and $5.5$, and at the helium abundance $\log{N({\rm He})/N({\rm H})}=-0.5, 0.0$ and $0.5$. We fitted the Balmer line spectrum (from H$\alpha$ to H9) and the \ion{He}{1} line spectrum ($\lambda\lambda$4026, 4471, 4713, 4921, and 5015\AA)  of CPD$-$20~1123 using $\chi^2$ minimization techniques. Table~\ref{tbl2} lists the atmospheric parameters $T_{\rm eff}$, $\log{g}$, $\log{N({\rm He})/N({\rm H})}$, and $M_V$ (see below) of CPD$-$20~1123 from this study, and the parameters of PG~0229$+$064 (note that $n_{\rm He}=N({\rm He})/[N({\rm H})+N({\rm He})]$) from \cite{ahm2003}. We also estimated the absolute magnitude $M_V$ of PG~0229$+$064 using parameters from \cite{ahm2003}. The helium to hydrogen abundance ratio is a factor of $\approx7$ larger in CPD$-$20~1123 than in PG~0229$+$064, and it is comparable to abundance ratios measured in a few He-sdB stars studied by \cite{ahm2003}. 

\begin{deluxetable*}{lccccc}
\tablewidth{0pt}
\tablecaption{Stellar Parameters \label{tbl2}}
\tablehead{
\colhead{Star\tablenotemark{a}}  &  \colhead{$T_{\rm eff}$} & \colhead{$\log{g}$} &  \colhead{$M_V$} & \colhead{$\log{N({\rm He})/N({\rm H})}$} & \colhead{$n_{\rm He}$}  \\
\colhead{} &  \colhead{($10^3$K)} & \colhead{} & \colhead{(mag)} & \colhead{} & \colhead{}}
\startdata
 CPD$-$20~1123  &  $19.8\pm0.4$ &  $4.55\pm0.10$ & $2.55\pm0.24$ & $0.15\pm0.15$ & $0.6\pm0.1$ \\
 PG~0229$+$064  &  $19.70\pm0.15$   &  $4.2\pm0.1$ & $1.70\pm0.24$ & $-0.66\pm0.03$ & $0.18\pm0.01$
\enddata
\tablenotetext{a}{Parameters for CPD$-$20~1123 from this study and parameters for PG~0229$+$064 from \citet{ahm2003}.}
\end{deluxetable*}

The stellar parameters of CPD$-$20~1123 place it on a extreme horizontal branch track at $M=0.49$ M$_\odot$ \citep{dor1993}. We obtained the evolutionary tracks at $M=0.471$, $0.480$, and $0.490$ M$_\odot$ with a solar abundance ($Y=0.288$, [Fe$/$H]$=$[O$/$Fe]=0.0) using the VizieR database at the Centre de Donn\'ees de Strasbourg. We then located CPD$-$20~1123 near the track at $M=0.49$ M$_\odot$  in the $\log{g}$ versus $T_{\rm eff}$ plane. Adopting a gravity of $\log{g}=4.55$, this mass implies a radius of $0.62\pm0.07$ R$_\odot$. Next we calculated the absolute $V$ magnitude using the solid angle subtended by the star at 10 pc, $\Omega = 2\times10^{-18}$, and the model flux at 5500 \AA.
Based on the apparent optical magnitude from the All Sky Automated Survey  \citep[ASAS-3;][]{poj2002} $V=12.08\pm0.04$,  and the calculated absolute magnitude $M_V=2.55\pm0.24$, we estimate a distance of $d=800^{+100}_{-80}$ pc. Therefore, a proper motion  of 19 mas yr$^{-1}$ \citep{hog2000} corresponds to a tangential velocity of $72^{+9}_{-7}$ km s$^{-1}$ characteristic of an old population in the thin disk or possibly in the thick disk \citep{the1997,alt2004}. A detailed  kinematical study awaits high-dispersion spectroscopy and accurate radial velocity measurements . On the other hand the location of CPD$-$20~1123 in the $\log{g}$ versus $T_{\rm eff}$ plane also sits well with the He-He merger tracks of \cite{sai2000}  displayed by \cite{ahm2003}.

The 2MASS measurements  do not support the presence of a cool companion to CPD$-$20~1123 \citep{cab2007}. Infrared excess is a common feature among subdwarf stars and has often been used to infer the presence of a cool companion \citep{the1995}. Although the \ion{C}{2} $\lambda4267$\AA\ line is clearly present in PG~0229$+$064, it is not detected in CPD$-$20~1123. The detection of weaker lines  will be best achieved in the future with the acquisition of high-dispersion spectra. Abundance of trace elements offers clues to the origin of these objects. In the present instance, the absence of carbon would support a He-He white dwarf merger origin. Overall, the origin of CPD$-$20~1123 remains an open question.

\section{Conclusions}

We show that CPD$-$20~1123 is a He-sdB. The spectrum shows strong \ion{He}{1} lines along with the hydrogen Balmer line series. The atmosphere is dominated by helium, but high-dispersion spectra are required to carry-out a detailed abundance analysis. Present evidence indicates that CPD$-$20~1123 may be the result of He-He white dwarf merger, although post horizontal- or giant-branch evolutionary scenarios cannot be excluded.

\acknowledgements

A. K. is supported by GA \v{C}R 205/05/P186. S.V. acknowledges support from the College of Science at the Florida Institute of Technology. This paper makes use of data obtained from the Isaac Newton Group Archive which is maintained as part of the CASU Astronomical Data Centre at the Institute of Astronomy, Cambridge, and from the database VizieR at the Centre de Donn\'ees de Strasbourg. We thank the referee J.A. Caballero for a number of useful suggestions.

\end{document}